# The Cluster Distribution as a Test of Dark Matter Models. I: Clustering Properties

Stefano Borgani[1,2], Manolis Plionis[2], Peter Coles[3] and Lauro Moscardini[4]

[1] *INFN Sezione di Perugia, c/o Dipartimento di Fisica dell'Università, via A. Pascoli, I-06100 Perugia, Italy*
[2] *SISSA – International School for Advanced Studies, via Beirut 2-4, I-34013 Trieste, Italy*
[3] *Astronomy Unit, School of Mathematical Sciences, Queen Mary & Westfield College, Mile End Road, London E1 4NS, UK*
[4] *Dipartimento di Astronomia, Università di Padova, vicolo dell'Osservatorio 5, I-35122 Padova, Italy*

22 November 1994

**ABSTRACT**
We present extended simulations of the large-scale distribution of galaxy clusters in several different dark matter models, using an optimized version of the Zel'dovich approximation. The high accuracy and low computational cost of this technique allow us to run a large ensemble of fifty realizations for each model, and we are therefore able to make an accurate determination of the cosmic variance. Six different dark matter models are studied in this work: Standard CDM, Open CDM ($\Omega_\circ = 0.2$), Tilted CDM with primordial spectral index $n = 0.7$, mixed Cold + Hot dark matter with $\Omega_{\rm hot} = 0.3$, spatially flat low-density CDM with cosmological constant term corresponding to $\Omega_\Lambda = 0.8$ and low Hubble constant ($h = 0.3$) CDM. We compare the discrete and smoothed cluster distributions with those of an Abell/ACO redshift sample using various statistical quantities, such as the $J_3$ integral and the probability density function (pdf). We find that the models that best reproduce the behaviour of $J_3(R)$ (i.e. the integral of the two-point correlation function), the pdf and the moments of the Abell/ACO cluster sample are the Cold + Hot dark matter model and a low-density CDM model with a non-zero cosmological constant: all the other models are ruled out at a high confidence level. The pdfs of all models are well approximated by a lognormal distribution, even when the Gaussian smoothing radius is as large as 40 $h^{-1}$ Mpc, consistent with the findings for Abell/ACO clusters. The low-order moments of all the pdf's are found to obey a variance-skewness relation for the form $\gamma \approx S_3 \sigma^4$, with $S_3 \simeq 1.9$, independent of the primordial spectrum shape and consistent with the observational results. Comparing the variance of the smoothed cluster density field to that of the dark matter distribution, we find that the linear biasing parameter for the simulated clusters is nearly constant over a large range of scales, but its value depends on the model. This suggests that it is probably a reliable procedure to use linear biasing to infer the dark matter power-spectrum from observational cluster samples. We also note that the abundances of clusters predicted using the Press-Schechter theory provide strong constraints on these models: only the mixed Cold + Hot Dark Matter model and the low-$H_\circ$ CDM model appear to produce the correct number-density of clusters. Taking this constraint together with the cluster clustering statistics leads one to conclude that the best of our models is the Cold + Hot dark matter scenario.

## 1 INTRODUCTION

The study of the distribution of matter on the largest scales provides important constraints on models of cosmic structure formation. If the gravitational instability picture is correct, the expected displacements of clusters of galaxies away from their primordial positions are much smaller than the typical separation of these objects. In principle, therefore, clusters of galaxies can yield clues about the primordial spectrum of perturba-tions that gave rise to them. This is the reason why so much effort has been devoted to compiling deep cluster surveys, starting with the the pioneering work of Abell (1958), Zwicky et al. (1968) and Abell, Corwin & Olowin (1989), and leading up to extended redshift surveys both in the optical (e.g. Postman, Huchra & Geller 1992; Dalton et al. 1994; Collins et al. 1994, and references therein) and in the X-ray (e.g. Nichol, Briel & Henry 1994; Romer et al. 1994) regions of the spectrum.

Accompanying the observational challenge of acquiring extended cluster redshift surveys, a great deal of effort has also been directed towards the provision of reliable statistical characterizations of the cluster distribution. It has been established that the cluster two-point correlation function is well modelled by a power-law,

$$\xi(r) = (r/r_\circ)^{-\gamma}. \qquad (1)$$

Although the slope, $\gamma \simeq 1.8$, turns out to be quite similar to that of galaxies, the correlation length, $r_o$, is much larger (cf. Bahcall & Soneira 1983; Klypin & Kopylov 1983; Bahcall 1988). Different determinations, based on different cluster samples, indicate values in the range $r_\circ = 13\text{--}25\,h^{-1}\,\text{Mpc}^\star$ (e.g., Nichol et al. 1994, and references therein), while more recently the reliability of the power-law model for $\xi(r)$ has also been questioned by different authors (cf. Olivier et al. 1993).

In order to compare the observational cluster data sets with different cosmological models several authors have resorted to large N-body simulations which were designed to sample the length-scales relevant to the cluster distribution (e.g. White et al. 1987; Bahcall & Cen 1992; Croft & Efstathiou 1994). The problem with this kind of approach is that large N-body simulations are very expensive from a computational point of view. Therefore, one is usually forced to consider only a limited number of models, with a small number of independent realizations for each model. In this respect, analytical approaches, based either on Eulerian linear theory (e.g. Bardeen et al. 1986; Coles 1989; Lumsden, Heavens & Peacock 1989; Borgani 1990; Holtzman & Primack 1993) or on the Zel'dovich approximation (e.g. Doroshkevich & Shandarin 1978; Mann, Heavens & Peacock 1993), are in general preferred to numerical simulations. Nevertheless, they are of limited utility since one is often obliged to resort to oversimplifying assumptions about the nature of galaxy clusters. Furthermore, statistics which go beyond the two-point correlation function and its Fourier transform, the power-spectrum, are hard to handle. Finally, it is not clear how shot-noise effects and/or observational biases (e.g., redshift-space distortions, selection functions, non-trivial sample geometry) can be realistically modelled in order to allow a consistent comparison with real data sets.

---

$\star$  $h$ is the Hubble constant in units of 100 km s$^{-1}$ Mpc$^{-1}$.

In a previous paper (Borgani, Coles & Moscardini 1994, hereafter Paper I) we used the Zel'dovich approximation (ZA hereafter) to generate cluster simulations which were accurate when compared to N-body simulations, and, at the same time, computationally cheap so as to enable us to produce many realizations of several dark matter (DM) models (see also Blumenthal, Dekel & Primack 1988). Recently, Sathyaprakash et al. (1994) compared the ZA and several approximations of non-linear gravitational clustering to direct N-body results. They showed that, although the ZA fails to follow small-scale clustering in the multistream region, it is nevertheless able to account for non-local effects due to long wavelength modes in the density fluctuation spectrum. For this reason, the ZA is not expected to provide a correct description of the internal structure and mass distribution of non-linear structures like galaxy clusters, but it is very accurate in locating them in the correct positions and thus reliably describes their spatial distribution. In this respect, it is not necessary to employ the full power and sophistication of modern N-body methods to investigate cluster clustering on large scales. On the other hand, accounting for the details of non-linear gravitational clustering becomes crucial if one is interested in small-scale aspects of cluster properties.

In this paper we present cluster simulations, which are based on an implementation of the ZA, that has been substantially improved with respect to that used in Paper I (see also Plionis et al. 1994, hereafter Paper II). We will show that these simulations provide as accurate a picture of the cluster distribution as can be obtained using N-body experiments. With no significant computation cost we generate 50 realizations for each of six different initial power-spectra we will consider and analyze the cluster simulations using two different statistical methods, described further below.

The availability of such a large set of simulations represents an extremely powerful tool to put constraints on DM models through a detailed comparison of the statistical properties of the real and simulated cluster distributions. However, a reliable quantitative measure of cluster clustering is not easy to find. Clusters are rather rare objects with typical mean separation of several tens of Mpcs; while bright galaxies have a mean separation comparable to their correlation length, clusters have a mean separation which is twice the corresponding $r_\circ$ value. For this reason, shot-noise effects become important on small scales ($\lesssim 10\,h^{-1}\,\text{Mpc}$) while, on larger scales ($\gtrsim 40\,h^{-1}\,\text{Mpc}$), a low signal-to-noise ratio is expected because the clustering is weak. Robust statistical estimators, which are able to provide reliable measures over a large range of scales, are required to describe the cluster distribution properly and to allow an effective comparison with model predictions.

As a first statistical test, we will use in this paper the quantity $J_3(R)$, which is defined through the integral of $\xi(r)$:

$$J_3(R) = \frac{1}{4\pi} \int_0^R \xi(r) r^2 dr. \qquad (2)$$

The advantage of using $J_3(R)$ over $\xi(r)$ lies on the fact that in a sparse distribution of objects, an integral quantity such as that defined by eq. (2), should be less susceptible to statistical noise than a differential quantity, such as $\xi(r)$.

An alternative method, which is becoming increasingly popular, is the study of the probability density function (pdf) itself. Usually one attempts to obtain a continuous density field by smoothing the discrete distribution of objects with some window function (a top-hat or a Gaussian one are the most commonly used). The smoothing procedure itself reduces significantly the shot–noise, which could dominate the discrete distribution (Gaztañaga & Yokoyama 1993). Then one can define the pdf $f(\varrho)$, where $\varrho = \rho/\langle\rho\rangle$, and derive its moments, defined by:

$$\langle \delta^n \rangle = \int_{-1}^{\infty} \delta^n f(\delta) d\delta, \qquad (3)$$

where $\delta = \varrho - 1$.

The pdf and moments of different galaxy samples have been estimated by various authors (e.g. Saunders et al. 1991; Bouchet et al. 1993; Gaztañaga and Yokoyama 1993; Sheth, Mo & Saslaw 1994; Gaztañaga 1994). Kofman et al. (1994) have compared the pdf derived from CDM N–body simulations with that of the IRAS sample and the one recovered using the POTENT procedure with $\Omega = 1$ (see also Lahav et al. 1993). Their main conclusion is that, if galaxies trace the mass, the observed pdf is consistent with Gaussian initial conditions.

Plionis & Valdarnini (1994, hereafter PV94) studied the pdf (and its moments) of the 3–D Abell/ACO smoothed cluster distribution and compared them with static simulations, based on a Gaussian fluctuation spectrum, which reproduced the two– and three–point cluster correlation functions as well as the observed selection effects. They found that the real and simulated cluster pdf is well approximated by a lognormal distribution. Cappi & Maurogordato (1994) have realized a study of the higher–order moments for the discrete Abell/ACO cluster distribution, for both projected and redshift samples, while Kolatt, Dekel & Primack (1994) estimated the pdf for real cluster samples as well as for cluster N–body simulations based both on Gaussian and non–Gaussian initial CDM fluctuations. They concluded that no evidence of non–Gaussian initial conditions are imprinted into the shape of the cluster pdf. In Paper II we compared the variance and the skewness of the smoothed Abell/ACO cluster pdf with those obtained from the ZA simulations of a list of DM models.

In this paper we will derive the pdf statistics of our cluster simulations. Using the same analysis procedure as that used for the Abell/ACO cluster sample (PV94), we will be able to put stringent constraints on the models we consider.

The layout of this paper is as follows. In Section 2 we describe the cluster simulations, i.e., how to optimize the ZA, the method of cluster identification and the considered models for the power–spectrum. In view of the unsuitability of our simulation method for studying the small–scale structure of clusters, we also use the Press & Schechter (1974; PS hereafter) method to compare the cluster abundances predicted by DM models with the available observational data. This analysis provides an independent constraint on the models we are considering. In Section 3 we describe the Abell/ACO sample we use. In Section 4 we present the analysis of the discrete cluster distribution using the $J_3$ integral, while in Section 5 we present the pdf and moment analysis of the smoothed cluster distribution. In Section 6 we discuss our results and state our main conclusions.

## 2 THE SIMULATIONS

### 2.1 The Zel'dovich approach

The Zel'dovich approximation (Zel'dovich 1970; Shandarin & Zel'dovich 1989) is based on the assumption of laminar flow for the motion of a self–gravitating non–relativistic collisionless fluid. Let $\mathbf{q}$ be the initial (Eulerian) position of a fluid element and $\mathbf{r}(\mathbf{q},t) = a(t)\,\mathbf{x}(\mathbf{q},t)$ the final position at the time $t$, which is related to the comoving Lagrangian coordinate $\mathbf{x}(\mathbf{q},t)$ through the cosmic expansion factor $a(t)$. The ZA amounts to assume the expression

$$\mathbf{r}(\mathbf{q},t) = a(t)\left[\mathbf{q} + b(t)\nabla_{\mathbf{q}}\psi(\mathbf{q})\right] \qquad (4)$$

for the Eulerian–to–Lagrangian coordinate mapping. In eq.(4) $b(t)$ is the growing mode for the evolution of linear density perturbations and $\psi(\mathbf{q})$ is the gravitational potential, which is related to the initial density fluctuation field, $\delta(\mathbf{q})$, through the Poisson equation

$$\nabla^2 \psi(\mathbf{q}) = -\frac{\delta(\mathbf{q})}{a(t)}. \qquad (5)$$

As a result of the factorization of the $t-$ and $\mathbf{q}-$ dependence in the displacement term of eq.(4), the fluid particles move under the ZA along straight lines, with comoving peculiar velocity

$$\mathbf{v}(\mathbf{q},t) = \dot{\mathbf{x}}(\mathbf{q},t) = \dot{b}(t)\nabla_{\mathbf{q}}\psi(\mathbf{q}). \qquad (6)$$

Therefore, gravity determines the initial kick to the fluid particles through eqs.(5) and (6), and afterwards they do not feel any tidal interactions. Particles fall inside gravitational wells to form structures, which however quickly evaporate. In this sense, the ZA gives a good description of gravitational dynamics as far as particle trajectories do not intersect with each other, while its validity breaks down when shell–crossing occurs, and local gravity dominates.

Several prescriptions have been suggested to overcome the shortcomings of the ZA, such as adding a small viscous term to the equation of motion for the fluid, or by going to higher–orders in Lagrangian perturbative theory (e.g. Sahni & Coles 1994, and references therein). As a further possibility, Coles, Melott & Shandarin (1993) have shown that filtering out the small–scale wavelength modes in the linear power–spectrum reduces the amount of shell–crossing, thus improving the performance of the ZA. Melott, Pellman & Shandarin (1993) claimed that an optimal filtering procedure is obtained by convolving the linear power–spectrum with the Gaussian filter

$$W_G(kR_f) = e^{-(kR_f)^2/2} . \qquad (7)$$

The problem then arises of suitably choosing the filtering radius $R_f$, in order to suppress shell–crossing as much as possible, without however preventing genuine clustering to build up. In Paper I we chose $R_f$ so that the expected mass within a Gaussian window of that radius were of the same order ($\sim 10^{15} M_\odot$) of the mass for a rich galaxy cluster. The disadvantage of this approach is that it does not rely on any objective criterion to optimize the ZA and treats in the same fashion different fluctuation spectra, which should produce a different amount of shell–crossing. The resulting filtering radius, $R_f = 5\Omega_\circ^{-1/3} h^{-1}$ Mpc, is generally larger than the optimal ones, which we use in the present paper, thus causing an excessive removal of clustering.

Kofman et al. (1994) derived the analytical expression for the average number of streams at each Eulerian point, $N_s$, as a function of the r.m.s. fluctuation level of the initial Gaussian density field. In Figure 1 we plot $N_s$ as a function of the filtering scale $R_f$ for the six different power–spectra that we will consider (see next subsection), evaluated according to eq.(7) of Kofman et al. (1994). As a general criterion, we decided to choose $R_f$ for each model so that $N_s = 1.1$. We found this to be a reasonable compromise between smaller $N_s$ values, giving rapidly increasing $R_f$ and high suppression of clustering, and larger $N_s$, at which the ZA progressively breaks down. The resulting r.m.s. fluctuation value corresponding to $N_s = 1.1$ is $\sigma = 0.88$.

By adopting this implementation of the ZA, the main steps of our cluster simulations are the following:

(a) Convolve the linear power–spectrum with the Gaussian window of eq. (7), and $R_f$ chosen as previously described.
(b) Generate a random–phase realization of the density field on $128^3$ grid points for a cubic box of $L = 320 h^{-1}$ Mpc aside.
(c) Move $128^3$ particles having initial Lagrangian position on the grid, according to the ZA. Each particle carries a mass of $4.4 \times 10^{12} h^{-1} \Omega_\circ M_\odot$.

(d) Reassign the density and the velocity field on the grid through a TSC interpolation scheme (e.g. Hockney & Eastwood 1981) for the mass and the moment carried by each particle.
(e) Select clusters as local density maxima on the grid according to the following prescription. If $d_{cl}$ is the average cluster separation, then we select $N_{cl} = (L/d_{cl})^3$ clusters as the $N_{cl}$ highest density peaks. In the following, we assume $d_{cl} = 40 h^{-1}$ Mpc, which is appropriate for the combined Abell/ACO cluster sample to which we will compare our simulation results (see Section 3). Therefore, we will analyze a distribution of 512 clusters in each simulation box, with periodic boundary conditions.

### 2.2 Dark matter models

We run simulations for six different models of the initial fluctuation spectrum. For each model, we generate 50 random realizations, so as to reliably estimate the effect of cosmic variance. All the models, except the open CDM one, are normalized to be consistent with the COBE measured quadrupole of CMB temperature anisotropy (Bennett et al. 1994).

The models we have considered are the following.

(1) The standard CDM model (SCDM), with $\sigma_8 = 1$ for the r.m.s. fluctuation amplitude within a top–hat sphere of $8 h^{-1}$ Mpc.
(2) A tilted CDM model (TCDM), with $n = 0.7$ for the primordial spectral index. Tilting the primordial spectral shape from the scale–free one has been suggested in order to improve the CDM description of the large–scale structure (e.g. Cen et al. 1992; Tormen et al. 1993; Liddle & Lyth 1993; Adams et al. 1993; Moscardini et al. 1994).
(3) A low Hubble constant CDM model (LOWH), with $h = 0.3$. Decreasing the Hubble constant has the effect of increasing the horizon size at the equivalence epoch, thus pushing to larger scales the falloff of the spectrum to the scale–free shape. The relevance of this models in alleviating several cosmological problems has been recently emphasized by Bartlett et al. (1994).
(4) A Cold + Hot DM model (CHDM), with $\Omega_{\rm hot} = 0.3$ for the fractional density contributed by the hot particles. For a fixed large–scale normalization, adding a hot component has the effect of suppressing the power–spectrum amplitude at small wavelengths (see, e.g. Klypin et al. 1993, and references therein, for the relevance of CHDM). Although the small–scale peculiar velocities are lowered to an adequate level, the corresponding galaxy formation time is delayed so that such a model is strongly constrained by the detection of high–redshift objects (e.g. Klypin et al. 1994; Ma & Bertschinger 1994, and references therein).

Table 1. The models. Column 2: the density parameter $\Omega_0$; Column 3: the cosmological constant term $\Omega_\Lambda$; Column 4: the density parameter of the hot component $\Omega_{hot}$; Column 5: the primordial spectral index $n$; Column 6: the Hubble parameter $h$; Column 7: the linear r.m.s. fluctuation amplitude at $8\,h^{-1}$ Mpc $\sigma_8$.

| Model | $\Omega_0$ | $\Omega_\Lambda$ | $\Omega_{hot}$ | $n$ | $h$ | $\sigma_8^{-1}$ | $R_f$ |
|---|---|---|---|---|---|---|---|
| SCDM | 1.0 | 0.0 | 0.0 | 1.0 | 0.5 | 1.0 | 4.4 |
| TCDM | 1.0 | 0.0 | 0.0 | 0.7 | 0.5 | 2.0 | 1.6 |
| LOWH | 1.0 | 0.0 | 0.0 | 1.0 | 0.3 | 1.6 | 2.4 |
| CHDM | 1.0 | 0.0 | 0.3 | 1.0 | 0.5 | 1.5 | 2.2 |
| OCDM | 0.2 | 0.0 | 0.0 | 1.0 | 1.0 | 1.0 | 4.5 |
| ΛCDM | 0.2 | 0.8 | 0.0 | 1.0 | 1.0 | 1.3 | 3.3 |

(5) An open CDM model (OCDM), with $\Omega = 0.2$. Lowering the density parameter has the effect of increasing the horizon size at the matter/radiation equivalence time, so as to shift the peak of the power–spectrum at larger scales.

(6) A spatially flat, low–density CDM model (ΛCDM), with $\Omega_o = 0.2$, $\Omega_\Lambda = 0.8$ for the cosmological constant term. While neglecting the spatial curvature, as required by inflation, this model has been shown to safely describe the large–scale distribution of galaxies (Baugh & Efstathiou 1993; Peacock & Dodds 1994) and galaxy clusters (Bahcall & Cen 1992; Croft & Efstathiou 1994).

The transfer functions for the above models have been taken from Holtzman (1989), except that of LOWH, which is taken from Bond & Efstathiou (1984), with suitably chosen shape parameter $\Gamma = \Omega_o h = 0.3$. All the model parameters are listed in Table 1.

Each power–spectrum is suitably smoothed on the scale $R_f$ according to the prescription described in Section 2.1. In Figure 1 we plot the average stream number per Eulerian point, $N_s$, as a function of the filtering radius for all the above models. The intersection of the $N_s = 1.1$ line with each curve indicates the smoothing scale adopted for the corresponding spectrum (see also Table 1). Note that the larger small–scale power for SCDM and OCDM requires a stronger filtering to suppress shell–crossing.

### 2.3 Cluster Abundances

As an independent constraint on the above models, we have computed the expected cluster abundances, as predicted by the standard PS formalism. If structures are identified through a filter $W$ on a scale $R$, so as to have a mass $M = f\bar{\rho}R^3$ ($\bar{\rho}$ is the average matter density), the PS formula for the number density of objects with mass between $M$ and $M + dM$ is

$$n(M)\,dM = \frac{1}{\sqrt{2\pi}}\frac{\delta_c}{f}\int_R^\infty \frac{\eta(R)}{\sigma(R)}\exp\left(-\frac{\delta_c^2}{2\sigma^2(R)}\right)\frac{dR}{R^2}, \quad (8)$$

where

$$\eta(R) = \frac{1}{2\pi^2\sigma^2(R)}\int k^4\, P(k)\,\frac{dW^2(kR)}{d(kR)}\,\frac{dk}{kR};$$

$$\sigma^2(R) = \frac{1}{2\pi^2}\int k^2\, P(k)\, W^2(kR)\, dk. \quad (9)$$

Therefore, the total abundance of objects of mass larger than $M$ is

$$N(>M) = \int_M^\infty n(M')\,dM'. \quad (10)$$

In the above expressions $f$ is a "form factor", which depends on the shape of the filter $W$: $f = (2\pi)^{3/2}$ for a Gaussian filter, and $f = 4\pi/3$ for a top–hat filter. A Gaussian filter will be assumed in the following analysis. The parameter $\delta_c$ is the critical density contrast, which represents the threshold value for a fluctuation to turn into an observable object, if evolved to the present time by linear theory. Arguments based on a simple spherical collapse suggest $\delta_c = 1.68$, but the inclusion of non–linear effects, as well as aspherical collapse, may lead to a lower value of $\delta_c$. For example, Klypin & Rhee (1994; KR94 hereafter) found that the cluster mass function in their CHDM N–body simulations is well fit by eq.(8) by taking $\delta_c = 1.5$.

White, Efstathiou & Frenk (1993) resorted to $X$–ray data for the temperatures of the gas component of clusters and estimated a cluster abundance of about $4 \times 10^{-6}(h^{-1}\,\mathrm{Mpc})^{-3}$ for masses exceeding $M = 4.2 \times 10^{14}M_\odot$. Using observed cluster velocity dispersions, Biviano et al. (1993) obtained an abundance of about $6 \times 10^{-6}(h^{-1}\,\mathrm{Mpc})^{-3}$ for clusters exceeding the above mass limit.

In Figure 2 we compare model predictions at different $\delta_c$ values to the above observational estimates. Note that realistic uncertainties on cluster abundances are probably larger than the difference between the two above values. They should include variations in the average cluster number density between different samples, biases toward high mass for observations of cluster velocity dispersions, uncertainties in the model used to relate gas temperature and cluster mass, etc. Although taking in mind such warnings, we note from Figure 2 that both low–density models are ruled out, with ΛCDM producing with $\delta_c = 1.5$ more than one order of magnitude less clusters than observed. This agrees with the suggestion of White et al. (1993), that a higher normalization ($\sigma_8 \simeq 1.4$) is required for these models to produce a correct number of clusters. In the $\Omega_o = 1$ models, it turns out that the resulting abundances depend mostly on the $\sigma_8$ normalization value and not on the shape of the spectrum. This is not surprising: the Gaussian smoothing scale, $R \simeq 4.2\,h^{-1}$Mpc, which encompasses a mass of $4\times 10^{-6}(h^{-1}\,\mathrm{Mpc})^{-3}$, is equivalent to a top–hat sphere of about $7\,h^{-1}$ Mpc, which is rather close to the normalization scale. As a result, SCDM turns out to produce too many clusters for any reasonable value of $\delta_c$. On the other hand, the low normalization of TCDM turns into a severe underproduction of clusters, even at the smallest $\delta_c$ values. The only two models which generate cluster abundances in agreement with the results by White et al. (1993) and Biviano et al. (1993) for a reasonable choice of $\delta_c$ are LOWH and CHDM.

It is wise to sound a note of caution about the strength of the constraints emerging from the Press–Schechter analysis, for a number of reasons. First, there is some evidence of a discrepancy between the mass profiles of clusters inferred from X–ray data (e.g. Edge & Stewart 1991) and from gravitational lensing considerations (e.g. Kaiser et al. 1994), so it is not clear whether the cluster mass function inferred from the X–ray data is correct. Likewise, there is a possibility that the distribution of cluster peculiar velocities may be affected by subclustering. One should also mention that the applicability of the Press-Schechter method is itself open to some doubt. Although, as we mentioned above, it appears to perform well for the CHDM model when compared with $N$–body simulations, its accuracy is yet to be verified for the other models. In the case of open CDM, where a much higher fraction of the cluster mass is baryonic than in the other models, one might imagine this formalism to be particularly suspect. We therefore take the constraints emerging from this analysis to be indicative but not watertight.

### 2.4 Reliability of the ZA

Before entering into the presentation of our analysis, we want to stress once more the reliability of the ZA for simulating the large–scale distribution of galaxy clusters. In Figure 3 we plot the projected particle distribution within a slice $10 \, h^{-1}$ Mpc thick, superimposing the cluster distribution, for the SCDM and OCDM models. In order to better show how the identified clusters trace the underlying density field, we used simulations within a $640 \, h^{-1}$ Mpc box[†]. The same initial phases have been used in both models, so that the structures present in the two slices can be directly compared. As expected, clusters are preferentially located at the knots corresponding to the intersections between filaments, while they avoid long filaments and flattened pancakes. As for the difference between the two models, it is apparent that the cluster distribution in the OCDM generates longer cluster filaments, surrounding larger underdense regions than in the corresponding SCDM model. Furthermore, in the OCDM simulation there are regions of size $\sim 150 \, h^{-1}$ Mpc which are completely devoid of clus-ters, while in the SCDM the clusters appear to be more *space–filling*.

In Figure 4 we compare the two–point cluster correlation function for our CHDM simulations (open dots) to that obtained by KR94 by evolving the same spectrum with a Particle–Mesh (PM) N-body code. Our results refer to the average taken over 50 random realizations. Error bars are estimated as the r.m.s. scatter over this ensemble. The KR94 results are obtained as an average over 2 realizations and error bars are quasi–Poissonian sampling uncertainties. By comparing this plot with Fig. 2 of Paper I, it is apparent that we have improved our implementation of the ZA by increasing the resolution and by optimizing the power–spectrum filtering. The agreement between ZA and N–body results is really remarkable and extends down to quite small scales ($\simeq 7 \, h^{-1}$ Mpc), where shell–crossing should already play some role. This result further confirms the reliability of the ZA to follow correctly the mildly non-linear clustering regime and ensures that our simulations provide a fair representation of the cluster distribution.

## 3 THE CLUSTER SAMPLE

We use the combined Abell/ACO $R \geq 0$ cluster sample, as defined in Plionis & Valdarnini (1991) [hereafter PV91] and analysed in Plionis, Valdarnini & Jing (1992) [hereafter PVJ] and Plionis & Valdarnini (1994) [hereafter PV94]. The northern sample, with dec$\geq -17°$ (Abell), is defined by those clusters that have measured redshift $z \lesssim 0.1$, while the southern sample (ACO; Abell, Corwin & Olowin 1989), with dec$\leq -17°$, is defined by those clusters with $m_{10} \leq 16.4$ (note that with this definition and due to the availability of many new cluster redshifts only 7 ACO clusters have $m_{10}$ estimated redshifts from the $m_{10} - z$ relation derived in PV91). Both samples are limited in Galactic latitude by $|b| \geq 30°$. The redshifts have been taken from a number of studies, the references of which can be found in PV94. The total number of clusters in our samples is 357 and 157, for Abell and ACO respectively.

To take into account the effect of Galactic absorption, we assume the usual cosecant law:

$$P(|b|) = \mathrm{dex} \, [\alpha \, (1 - \csc |b|)] \qquad (11)$$

with $\alpha \approx 0.3$ for the Abell sample (Bahcall & Soneira 1983; Postman et al. 1989) and $\alpha \approx 0.2$ for the ACO sample (Batuski et al. 1989). The cluster–redshift selection function, $P(z)$, is determined in the usual way (cf. Postman et al. 1989; PVJ; PV94), by fitting the cluster density, as a function of $z$ (see the above reference for details). Cluster distances are estimated using the standard relation:

$$R = \frac{c}{H_\circ q_\circ^2 (1+z)} \left[ q_\circ z + (1 - q_\circ)(1 - \sqrt{2q_\circ z + 1}) \right] \qquad (12)$$

---

[†] The analysis of these simulations will be presented in forthcoming papers.

with $H_o = 100\ h$ km sec$^{-1}$ Mpc$^{-1}$ and $q_o = \Omega_o/2$. Strictly speaking, eq.(12) holds only for vanishing cosmological constant. Therefore, for a consistent comparison with the simulation models, we should use different $R$–$z$ relations for the Abell/ACO analysis. However, we verified that final results are essentially independent of the choice of the $(\Lambda, \Omega_o)$ parameters used in the simulations. For this reason, in the following we will present results for real data only based on assuming eq.(12) with $q_o = 0.2$.

PVJ and others have estimated the two–point correlation function for the Abell and ACO cluster samples and found that the slope of the two–point function has a value $\sim 1.8 \pm 0.2$ for both of them. Their amplitudes, however, are slightly different, with Abell clusters having $r_o \simeq 18 \pm 4\ h^{-1}$ Mpc (bootstrap errors) out to $\lesssim 50\ h^{-1}$ Mpc while the ACO clusters have $r_o \simeq 22 \pm 10\ h^{-1}$ Mpc but only out to $\sim 30\ h^{-1}$ Mpc; this fact could be attributed to the relatively small solid angle covered by the ACO and to the consequent undersampling of large wavelengths. Furthermore, based on the analysis of Jing, Plionis & Valdarnini (1992), we believe that the cluster correlations are not significantly affected by contamination effects (cf. Sutherland 1988).

PVJ and PV94 found that the Abell and ACO cluster number densities, out to their limit of completeness, are $\sim 1.4 \times 10^{-5}\ h^3$ Mpc$^{-3}$ and $\sim 2.1 \times 10^{-5}\ h^3$ Mpc$^{-3}$, corresponding to mean separations $d_{cl} \approx 41\ h^{-1}$ Mpc and $d_{cl} \approx 36\ h^{-1}$ Mpc, respectively. The higher space-density of ACO clusters is partly due to the huge Shapley concentration (Shapley 1930), but a significant part is also due to systematic density differences between the Abell and ACO cluster samples, as a function of $z$, which has been noted in a number of studies (cf. PV91 and references therein) and which could be attributed to the high sensitivity of the IIIa–J emulsion plates. In PV94, this effect was taken into account by normalizing the densities of the two samples using a radial *matching* function, $W(R)$, which is defined as the ratio between the average densities for Abell and ACO clusters at equal volume shells.

In the following, we compare results based on the Abell/ACO sample with those derived from our simulated cluster populations, selected so that $d_{cl} = 40\ h^{-1}$ Mpc. Variations in $d_{cl}$ of the order of the Abell-ACO difference, does not significantly affect the resulting statistical properties. Finally, following PV94 we restrict our analysis of the real cluster sample within a maximum distance of $R_{max} = 240\ h^{-1}$ Mpc, in order to minimize the uncertainties due to the approximate character of the redshift selection function, $P(z)$, and of the radial *matching* function, $W(R)$, especially at large distances.

## 4 STATISTICS OF THE DISCRETE CLUSTER DISTRIBUTION

Our first statistical test for comparing the real data and the simulations, involves the evaluation of the quantity $J_3(R)$, defined by eq. (2). It is straightforward from the definition of this quantity to construct the estimator

$$J_3(R) = \frac{R^3}{3}\left(\frac{N_{nb}}{\overline{N}} - 1\right), \qquad (13)$$

where $N_{nb}$ is the average number of cluster neighbours within a distance $R$ from a cluster, while $\overline{N}$ is the expected number of neighbours for a random cluster distribution (estimated at the positions of the real clusters). Therefore, $J_3(R) \propto R^{3-\gamma}$ as long as $\xi(r) \propto r^{-\gamma}$.

It has been argued (cf. KR94) that the scale at which the power–law shape of $\xi(r)$ breaks and firstly crosses zero, is a potentially powerful test for cosmological models. However, since such a scale corresponds by definition to the weak clustering regime, its detection can be heavily affected by statistical noise; for an explicit demonstration of this, see Paper I. In this respect, the analysis of $J_3$ should have the advantage of being more stable and suffering less from observational biases.

In Figure 5 we plot $J_3(R)$ for the real data (filled circles) and simulations (open circles). We estimated $\overline{N}$ for the real data by averaging over 100 random samples, having the same selection criteria (boundaries, galactic extinction function, redshift selection and systematic Abell/ACO differences) as the real one. Error bars for the simulated samples are $1\sigma$ scatter over the ensemble of 50 realizations. In Table 2 we report values of $J_3$ for data and simulations at three different scales. The quoted uncertainties for real cluster analysis are $1\sigma$ scatter estimated over an ensemble of 100 bootstrap resamplings. Such errors are not plotted in Figure 5. In fact, since we are asking which is the probability that a given model generates a result like that of the observed cluster distribution, its 'success' is just measured by the distance of the real data point from the cosmic r.m.s. error bars. This should be taken into account when judging to which confidence level a model has to be accepted or rejected. In Paper II we verified that intermediate scales of few tens of Mpcs are best suited to constrain DM models, when using the cluster distribution, smaller and larger scales being affected by shot-noise and low signal-to-noise ratio, respectively. In the present analysis we do not consider scales much smaller than $20\ h^{-1}$ Mpc as well as larger than $60\ h^{-1}$ Mpc.

For the Abell/ACO sample, $J_3(R)$ increases up to $R \simeq 35\ h^{-1}$ Mpc, flattens at a scale corresponding to the break of the power-law shape of $\xi(r)$, and eventually declines at $R \gtrsim 50\ h^{-1}$ Mpc, after which $\xi(r)$ becomes negative. By comparing this result with those of the simulations, it turns out that the only two models which overcome this test are CHDM and $\Lambda$CDM, although both of them seem to produce too strong clustering at the smallest scale considered and the second one is also marginally overclustered at the largest scale. All the other models are ruled out at least at a $3\sigma$ level. The cluster distributions for the SCDM, LOWH and TCDM models are too weakly clustered over the whole scale range. Note that the SCDM has a rather flat $J_3(R)$ profile, according to the expectation that this model has a cluster two–point correlation function which declines rapidly beyond $\sim 20\, h^{-1}$ Mpc. Conversely, OCDM generates too much clustering, with $J_3(R)$ increasing up to $R \gtrsim 60\, h^{-1}$ Mpc. These results confirm what we already found in Paper II from the analysis of the variance/skewness relation of the smoothed cluster density field. Note that the results for TCDM and LOWH are remarkably similar. This agrees with the expectation that, as far as the shape of the power–spectrum is concerned, a change in the Hubble parameter $h$ is roughly equivalent to a change in the spectral index $n$ according to the relation $\Delta h = -\Delta n$ (cf. Lyth & Liddle 1994).

**Table 2.** $J_3(R)$ values at different scales (in $h^{-1}$ Mpc units) for both simulated and real cluster distributions.

| Model | $J_3(R) \times 10^{-3} (h^{-1}\text{Mpc})^3$ | | |
|---|---|---|---|
| | 28.9 | 43.2 | 64.3 |
| SCDM | $5.00 \pm 0.64$ | $5.55 \pm 1.34$ | $3.86 \pm 2.12$ |
| TCDM | $4.93 \pm 0.74$ | $6.05 \pm 1.60$ | $5.71 \pm 2.74$ |
| LOWH | $5.12 \pm 0.79$ | $6.10 \pm 1.60$ | $5.16 \pm 2.83$ |
| CHDM | $8.39 \pm 0.94$ | $10.19 \pm 1.98$ | $8.77 \pm 3.36$ |
| OCDM | $11.89 \pm 1.73$ | $16.52 \pm 2.80$ | $18.66 \pm 6.05$ |
| $\Lambda$CDM | $8.43 \pm 1.03$ | $10.99 \pm 2.19$ | $11.61 \pm 4.27$ |
| Abell/ACO | $8.10 \pm 1.69$ | $10.28 \pm 2.53$ | $6.96 \pm 5.17$ |

## 5 STATISTICS OF THE SMOOTHED CLUSTER DISTRIBUTION

In order to facilitate the comparison of our results with those obtained by PV94 from the combined Abell/ACO cluster sample, we followed basically the same procedure as they did, and which we briefly describe below. We obtain a continuous cluster density field by smoothing the cluster distribution on a grid, with grid–cell width of $20\, h^{-1}$ Mpc ($16^3$ grid–points), using a Gaussian kernel:

$$\mathcal{W}(|\mathbf{x}_i - \mathbf{x_g}|) = \left(2\pi R_{sm}^2\right)^{-3/2} \exp\left(-\frac{|\mathbf{x}_i - \mathbf{x_g}|^2}{2R_{sm}^2}\right). \quad (14)$$

The smoothed cluster density, at the grid–cell positions $\mathbf{x_g}$, is then:

$$\rho(\mathbf{x_g}) = \frac{\sum_i \rho(\mathbf{x}_i)\mathcal{W}(|\mathbf{x}_i - \mathbf{x_g}|)}{\int \mathcal{W}(|\mathbf{x} - \mathbf{x_g}|)\mathrm{d}^3 x}, \quad (15)$$

where the sum is over the distribution of clusters with positions $\mathbf{x}_i$. In order to study the cluster density field at different smoothing scales, we use three radii for the Gaussian kernel: $R_{sm} = 20$, 30 and 40 $h^{-1}$ Mpc with $|\mathbf{x}_i - \mathbf{x_g}| \leq 3 R_{sm}$. Therefore the integral in the denominator of eq. (15) has a value smaller than unity ($\simeq 0.97$).

### 5.1 The probability density function

As a first test for the smoothed cluster density field, we work out the probability density function, $f(\varrho)$, which represents a low–order (one–point) statistics. We then compare the pdf of each set of cluster simulations with the observed Abell/ACO pdf, derived by PV94, as well as with the following theoretical models.

(a) The Gaussian distribution given by

$$f(\varrho) = \frac{1}{\sqrt{2\pi\sigma^2}} \exp\left[-\frac{(\varrho - 1)^2}{2\sigma^2}\right], \quad (16)$$

where $\sigma$ is the standard deviation of $\varrho$ ($\equiv \rho/\langle\rho\rangle$). If $f(\varrho)$ is a Gaussian then it should be defined in an infinite interval, which implies that $f(\varrho < 0) \neq 0$. Since, however $\varrho \geq 0$ by definition, $f(\varrho)$ is expected to be well approximated by a Gaussian only in the limit $\sigma \to 0$. In this case the skewness, $\gamma$ ($\equiv \langle \delta^3 \rangle$), vanishes. Even in the case of an initial Gaussian density field, the gravitational evolution acts in such a way as to increase the variance $\sigma^2$, and thus, due to the constraint $\varrho \geq 0$, $f(\varrho)$ has to become positively skewed. For as long as the variance $\sigma^2 \equiv \langle \delta^2 \rangle$ is small, the deviation of the pdf shape from a Gaussian is well approximated by the Edgeworth expansion (Colombi 1994).

(b) The lognormal distribution given by

$$f(\varrho) = \frac{1}{\sqrt{2\pi\sigma_L^2}} \exp\left[-\frac{(\ln \varrho - \mu_L)^2}{2\sigma_L^2}\right] \frac{1}{\varrho}, \quad (17)$$

where $\varrho$ is obtained through an exponential transformation of a Gaussian random variable $\chi$ as $\varrho = \exp(\chi)$. In eq. (17), $\mu_L$ and $\sigma_L$ are the mean and standard deviation of $\ln \varrho$ respectively. It has been argued that this distribution describes the distribution of density perturbations resulting from Gaussian initial conditions in the weakly non–linear regime Coles & Jones 1991). Bernardeau & Kofman (1994) have shown that the lognormal distribution is **not** really a natural consequence of mildly non–linear gravitational evolution, but a very convenient fit only in some portion of the $(\sigma, n)$–plane (i.e. $\sigma \ll 1$ and spectral index $n \approx -1$). It has nevertheless been found to give an extremely good fit to the CDM density and the IRAS galaxy pdf in the weakly–linear regime (Kofman et al. 1994), as well as to the observed Abell/ACO cluster distribution (PV94).

(c) The pdf following from the Zel'dovich approximation (Kofman et al. 1994):

$$f(\varrho) = \frac{9 \times 5^{3/2}}{4\pi N_s \varrho^3 \sigma^4}$$

$$\times \int_{3\varrho^{-1/3}}^{\infty} ds\, e^{-(s-3)^2/2\sigma^2} \left(1 + e^{-6s/\sigma^2}\right)$$

$$\times \left(e^{-\beta_1^2/2\sigma^2} + e^{-\beta_2^2/2\sigma^2} - e^{-\beta_3^2/2\sigma^2}\right);$$

$$\beta_n(s) = \sqrt{5}\,s\,\{1/2 + \cos[2/3\,(n-1)\pi + 1/3\,\arccos\left(54/\varrho s^3 - 1\right)]\},\qquad(18)$$

where $\sigma$ is the rms amplitude of density fluctuations and $N_s$ is the average stream number per Eulerian point.

Note that discreteness effects could be important since they affect the shape of the pdf and the estimation of its moments; especially at small $R_{sm}$, when the number of clusters in the Gaussian sphere is small, and/or when the smoothing fails to create a continuous density field due to discreteness (in our case this is apparent in the $R_{sm} = 20\,h^{-1}$ Mpc case for $\varrho \leq 0.8$). In the case of a Poisson sampling of an underlying continuous density field, the shot-noise contributions to the moments can be easily estimated and corrected for (cf. Peebles 1980). However, the cluster distribution can hardly be considered a Poissonian sampling of the underlying (galaxy) distribution, since clusters are expected to form at high density peaks. Therefore, the Poissonian shot-noise correction could not give a reasonable description of discreteness effects (Coles & Frenk 1991; Borgani et al. 1994). Moreover, Gaztañaga & Yokoyama (1993) have shown that the smoothing process itself considerably suppresses these shot-noise effects. For these reasons PV94 did not use any shot-noise corrections. To make a consistent comparison of our models with the data, we also did not include such corrections in our analysis. Since all the model cluster distributions have the same mean number density and we treat them similarly, the possible effects of shot-noise are accounted for in the same way in both the data and the simulations: we are therefore comparing like with like.

In Table 3 we present the results of the comparison between the simulation pdf, the PV94 Abell/ACO cluster pdf, the lognormal as well as the Gaussian distributions, using a $\chi^2$-test defined as:

$$\chi^2 = \sum_i^{bins} \left(\frac{f_i^{\text{sim}}(\varrho) - f_i^{\text{theor}}(\varrho)}{\epsilon_i}\right)^2 \qquad(19)$$

where the weights $\epsilon_i^2$ correspond to cosmic variance. Note that we derived the simulation pdf in redshift space so that a consistent comparison with the PV94 results can be made, while the comparison with theoretical models is done in real space.

In Figure 6 we present the simulation cluster pdfs for the most successful and least successful models (i.e. CHDM and SCDM), together with the PV94 Abell/ACO pdf at $R_{sm} = 20$ and $40\,h^{-1}$ Mpc, and in Figure 7 we make the comparison between these two cosmological models and the theoretical distributions. The error bars represent the scatter around the ensemble mean values (cosmic variance). As in Table 3, comparisons with theoretical models and real data are made in real space and in redshift space, respectively. There is an excellent agreement between the CHDM and the Abell/ACO cluster pdfs while in the SCDM case there is a clear discrepancy at small ($\leq 0.5$) and large $\varrho$'s.

**Table 3.** $\chi^2$ probabilities that the indicated simulation model pdf could have been drawn from a parent distribution given by the lognormal, Gaussian or the real cluster (PV94) pdf's. For the $R_{sm} = 20\,h^{-1}$ Mpc case the comparison is performed for $\varrho > 0.8$ (see text). Also reported in Column 6 are the values of the reduced skewness $S_3$ for real simulations and for real data.

| Model | $R_{sm}$ | $P_{\chi^2}^{LN}$ | $P_{\chi^2}^{G}$ | $P_{\chi^2}^{data}$ | $S_3$ |
|---|---|---|---|---|---|
| SCDM | 20 | $1.6\times10^{-2}$ | 0.00 | $6\times10^{-3}$ | $1.87\pm0.29$ |
|  | 30 | 0.83 | 0.00 | 0.75 | $1.83\pm0.70$ |
|  | 40 | 0.99 | 0.92 | 0.97 | $1.60\pm1.25$ |
| TCDM | 20 | 0.15 | 0.00 | $4\times10^{-4}$ | $1.96\pm0.32$ |
|  | 30 | 0.99 | 0.00 | 0.99 | $2.02\pm0.80$ |
|  | 40 | 0.99 | 0.06 | 0.99 | $1.96\pm1.44$ |
| LOWH | 20 | $1.6\times10^{-2}$ | 0.0 | $9\times10^{-4}$ | $1.87\pm0.28$ |
|  | 30 | 0.89 | 0.00 | 0.97 | $1.87\pm0.69$ |
|  | 40 | 0.99 | 0.49 | 0.97 | $1.75\pm1.18$ |
| CHDM | 20 | 0.10 | 0.00 | 0.95 | $1.93\pm0.30$ |
|  | 30 | 0.99 | 0.00 | 0.98 | $1.96\pm0.65$ |
|  | 40 | 0.99 | 0.00 | 0.63 | $1.81\pm1.08$ |
| OCDM | 20 | $3.8\times10^{-2}$ | 0.00 | 0.68 | $1.87\pm0.21$ |
|  | 30 | 0.94 | 0.00 | $7.5\times10^{-3}$ | $1.87\pm0.41$ |
|  | 40 | 0.99 | 0.00 | $1.1\times10^{-3}$ | $1.77\pm0.72$ |
| $\Lambda$CDM | 20 | 0.39 | 0.00 | 0.98 | $1.90\pm0.26$ |
|  | 30 | 0.33 | 0.00 | 0.72 | $1.96\pm0.55$ |
|  | 40 | 0.02 | 0.70 | 0.22 | $1.95\pm0.91$ |
| Abell/ACO | 20 | $2\times10^{-3}$ | 0.00 | – | $1.81\pm0.23$ |
|  | 30 | $6.3\times10^{-2}$ | 0.00 | – | $1.78\pm1.30$ |
|  | 40 | 0.99 | 0.27 | – | $1.76\pm1.85$ |

It is apparent that:

(i) The Gaussian distribution does not provide a good fit at any $R_{sm} \leq 30\,h^{-1}$ Mpc and for any model. For $R_{sm} = 40\,h^{-1}$ Mpc the Gaussian fit is acceptable only for the SCDM, LOWH and $\Lambda$CDM models.

(ii) The Zel'dovich pdf model is inconsistent with the simulation results, even though the underlying dynamics governing the cluster distribution are described by the ZA. We find that this distribution is ruled out at a confidence level larger than 99.99%, for $R_{sm} \leq 30\,h^{-1}$ Mpc for all the simulation models. This is the reason why we did not show results for this model in Table 3. The SCDM, TCDM, CHDM and OCDM models are only consistent at a $\sim 20\% - 25\%$ level for $R_{sm} = 40\,h^{-1}$ Mpc. One may argue that, since the ZA pdf is designed to describe the DM clustering, its failure for the cluster distribution is nothing but the consequence of not accounting for the mass within clusters in the analysis of their distribution. To check this, we repeated the analysis by weighting simulated clusters according to their mass and found no appreciable differences in the pdf shapes.

(iii) All simulation pdf's are well approximated by a lognormal distribution, irrespective of their different power–spectra. Note that at the $R_{sm} = 20\,h^{-1}$ Mpc case the comparison is done for $\varrho > 0.8$ because, at lower values of $\varrho$, discreteness effects introduce significant noise. Therefore, in contrast to the case of the matter distribution, the lognormal fit to the cluster pdf is more likely to be connected with the high–peak biasing description of cluster formation than being due to non–linear gravitational effects, which dominate much smaller scales (Bernardeau & Kofman 1994).

(iv) The scale which best discriminates between different models and the Abell/ACO data is clearly $R_{sm} = 20\,h^{-1}$ Mpc. At larger scales, all the models, except OCDM, produce acceptable fits. The only models that produce a pdf consistent with the PV94 results, at all 3 smoothing radii, are the CHDM and the $\Lambda$CDM, although CHDM fares better on larger scales, in agreement with the $J_3(R)$ results and with the results of Paper II.

### 5.2 Moments of the pdf

According to eq. (3), the moments of the pdf give a large weight to the high density tail ($\delta > 1$) of the pdf. They are therefore expected to suffer less from shot–noise effects, which are smaller in the overdense parts of the distribution. In Paper II we presented results about the variance, $\sigma^2 = \langle \delta^2 \rangle$, and the skewness, $\gamma = \langle \delta^3 \rangle$.

It has been argued on several grounds (e.g. Coles & Frenk 1991) that the relation

$$\gamma \approx S_3 \left(\sigma^2\right)^2, \qquad (20)$$

with $S_3$ nearly independent of scale, should describe the clustering of cosmic structures. Although at small scales, below a few Mpc, eq. (20) is predicted by models of non–linear gravitational clustering (e.g. Borgani 1994 and references therein), at the larger scales, sampled by galaxy clusters, it is expected to hold due to mildly non–linear evolution as well as by the bias relating the cluster and DM distributions. The resulting $S_3$ values at different $R_{sm}$ are reported in Table 3 for both the simulations and the Abell/ACO sample. We note that *(a)* the reduced skewness $S_3$ is always independent of the scale with a good accuracy, and *(b)* it takes the same value $S_3 \simeq 1.9$ for all the models, within statistical fluctuations, and consistent with the observational data (PV94).

Accordingly, we conclude that, for the cluster distribution, only the amplitude of clustering, and not its *nature*, depends on the initial power–spectrum. This suggests that both the lognormal pdf shape and the $S_3$ value observed for real data are natural consequences of high–peak biasing and possibly of the random–phase assumption of the primordial density field.

### 5.3 The bias parameter of clusters

It is interesting to attempt to derive from our simulations the cluster biasing parameter, which is defined as the ratio between the r.m.s. fluctuations of the cluster and DM distributions:

$$b_{cl} = \frac{\sigma_{cl}}{\sigma_{DM}}. \qquad (21)$$

The relevance of the biasing parameter lies in the fact that it is a necessary ingredient if one is seeking to derive the shape of the spectrum of the initial density fluctuations, starting from the distribution of clusters of galaxies. The usual assumption is that biasing is linear, that is $b_{cl}$ does not depend on the scale. From their power–spectrum analysis, Jing & Valdarnini (1993) and Peacock & Dodds (1994) verified that at least the relative biasing between clusters and optically as well as infrared selected galaxies is independent of the scale to a quite good accuracy. In particular, Peacock & Dodds (1994) found that $b_{cl}/b_{IRAS} \simeq 4.5$, for the relative biasing between clusters and IRAS galaxies. In our simulations, we can directly verify whether the linear biasing paradigm for clusters is reasonable, since we also know the variance of the linear DM fluctuations:

$$\sigma(R_{sm})_{DM} = \left[\frac{1}{2\pi^2}\int dk\,k^2\,P(k)\,\mathcal{W}_{R_{sm}}^2(k)\right]^{1/2}, \qquad (22)$$

where $\mathcal{W}_{R_{sm}}(k) = \exp(-k^2 R_{sm}^2/2)$ is the Fourier transform of the window function of eq. (14).

In Figure 8 we plot the biasing parameters for the cluster simulations at different $R_{sm}$. The heavy–dashed horizontal lines delineates the $1\sigma$ band for the observational result, once we assume that IRAS galaxies fairly trace the DM distribution, $b_{IRAS} = 1.0 \pm 0.2$. Independent of the model considered, $b_{cl}$ is fairly constant over the whole scale range. Figure 8 is quite different from the analogous plot presented in Paper I, which showed a decreasing trend for $b_{cl}$ at small scales. This confirms how important is the increased resolution and the optimization of the ZA for the reliability of our simulations. From one hand, such a linearity of the biasing is a rather remarkable result, since both the evolution of the den-sity field and the selection of clusters as high–density peaks represent definitely non–linear transformations of the initial fluctuations. From the other hand, this result supports the usual assumption of $b_{cl} = const$, used to infer the shape of the primordial power–spectrum from that of clusters. Note, however, that different models have rather different values of $b_{cl}$. The TCDM clusters are by far the most biased. The resulting $b_{cl} \simeq 7$ could be made consistent with observations only by allowing for a IRAS biasing as low as $b_{IRAS} \simeq 0.56$.

## 6 CONCLUSIONS

In this paper we have compared results of a statistical analyses of our simulated cluster distributions and of a combined sample of Abell/ACO clusters. Our cluster simulations, which are based on the Zel'dovich approximation, are extremely cheap computationally; each realization takes about 6 minutes of CPU on a HP755/125 workstation. This has allowed us to run a large number of realizations (50) for each model, so as to properly estimate the cosmic variance. We have also carefully verified that such a low computational cost is not at the expense of reliability. Our cluster selection procedure identifies clusters in correspondence of knots at the intersection of filaments (see Figure 3). Comparing the resulting two–point correlation function for our CHDM clusters with that obtained by Klypin & Rhee (1994) from their PM simulations of the same model, we find a extremely good agreement, even down to quite small scales ($\simeq 7\,h^{-1}$ Mpc), at which one can doubt the validity of the ZA. The reason for this remarkable success of the ZA in reproducing N–body results lies in its ability to account for non–local effects when moving particles from their initial (linear) positions to their correct evolved ones (cf. Pauls & Melott 1994; Sathyaprakash et al. 1994).

As a preliminary indication of the success of the DM models, we compared the predicted cluster abundances with available observational results using the Press & Schechter (1974) approach. The only models that produce adequate abundances are CHDM and low–$H_o$ CDM but, given the uncertainties in both theory and observation, we take these results to be indicative rather than definitive.

We have also analyzed both the point–like cluster distribution using the $J_3$ statistic, and the Gaussianly–smoothed cluster distribution. Both analyses provide consistent stringent constraints on the simulated DM models, which confirm the results presented in Paper II. The only models which pass all the tests are the Cold + Hot DM model and the low–density CDM model with non–vanishing cosmological constant (ΛCDM). Standard, Tilted and Low–$H_o$ CDM versions do not account for the large–scale clustering of the real cluster sample. The Open CDM model produces a much stronger clustering than real clusters on all scales.

We find that the shape of the probability density function (pdf) is not well approximated by a Gaussian distribution, even at the largest smoothing scale considered. The pdf for both simulations and Abell/ACO clusters is always much better reproduced by the lognormal model than by the Zel'dovich prediction, despite the fact that the ZA governs the underlying dynamics. This shows that, at least for clusters, the lognormal shape of the pdf does not occur by chance, for a limited set of initial conditions, as argued to happen for the galaxy distribution (Bernardeau & Kofman 1994). Instead it is much more likely to be related to the fact that clusters trace the high density peaks of the underlying matter field and, perhaps, to the initial random–phase assumption. Furthermore, using a $\chi^2$–test to compare the shape of the data and simulations pdf, we find that the best models are the CHDM and ΛCDM ones, in agreement with the $J_3$ and the moment analysis (Paper II).

Although the variance and skewness of the smoothed cluster pdf are powerful discriminators of different models (see Paper II), the reduced skewness, $S_3 = \gamma/\sigma^4$, turns out to be independent of the initial spectrum. We always find $S_3 \simeq 1.8$–$1.9$, almost independent of the scale, and consistent with the observational results (Plionis & Valdarnini 1994). One could be tempted to conclude that such a value of $S_3$ is naturally produced by the high–peak selection of clusters, probably combined with the Gaussian nature of the initial fluctuations. Whether the analysis of the reduced skewness represents a test of Gaussian vs. non–Gaussian initial conditions remains to be seen (see also Coles et al. 1993).

We have verified that the linear biasing prescription used to relate cluster and DM distributions is always satisfied to a good precision. The resulting value of the biasing parameter, $b_{cl}$, depends on the details of the model, so that it could represent a further discriminator between different initial power–spectra. Taking $b_{cl}/b_{IRAS} \simeq 4$–$5$ for the ratio between the biasing parameters for clusters and IRAS galaxies (cf. Peacock & Dodds 1994; Plionis 1994), and assuming that clusters trace the DM distribution fairly, we find that OCDM and LOWH models have a marginally too low and too high $b_{cl}$, respectively, while TCDM clusters are far too strongly biased tracers of the density field, compared to real clusters.

The overall picture emerging from these studies is that the large–scale cluster distribution places stringent constraints on models of structure formation. Observed cluster clustering is rather well reproduced either by the CHDM model with $\Omega_{hot} = 0.3$ or by a low density CDM model $\Omega_o = 0.2$, with non–vanishing cosmological constant term, $\Omega_\Lambda = 0.8$. If combined with results on the cluster abundances, this suggests that the only surviving model is CHDM, with ΛCDM producing more than one order of magnitude fewer clusters than observed. Furthermore, since ΛCDM behaves much like OCDM from the point of view of the large-scale velocity fields, our hope is that we should be able to further discriminate between CHDM and ΛCDM by analyzing the dipole structure of the cluster distribution (Tini Brunozzi et al. 1994) and cluster peculiar velocities.

One can also ask whether reasonable modifications of the parameters in the DM models considered above (i.e. the Cold + Hot DM mixture, the $\Omega_0$ and $\Lambda$ values, the primordial spectral index $n$, the Hubble constant, etc.) could lead to significant changes in the resulting cluster distribution. Verifying this will not be a difficult task, thanks to the low computational cost of our simulations, which makes them a flexible instrument to explore the entire parameter space of DM models. We believe that, in future investigations of specific DM models, the optimal strategy would be, first of all, to run optimal ZA simulations in order to assess the model on large scales. Only after that one should decide whether a model is worth exploring at smaller scales by means of high-resolution, computationally expensive N-body simulations.

### Acknowledgments.

MP acknowledges the receipt of an EC *Human Capital and Mobility* Fellowship. PC acknowledges the receipt of a PPARC Advanced Research Fellowship. SB and LM have been partially supported by Italian MURST. We are also grateful to PPARC for support under the QMW Visitors Programme in Astronomy GR/J 88357.

**Figure Captions**

**Figure 1.** The average number of streams per Eulerian point, $N_s$, as a function of the Gaussian filtering scale $R_f$ for the six different models. The dotted horizontal line, $N_s = 1.1$, delineates the quasi single-stream regime.

**Figure 2.** The abundances of clusters with mass $M > 4.2 \times 10^{14} M_\odot$ for the different DM models, as predicted by the Press & Schechter (1974) formalism, as a function of the critical density contrast $\delta_c$. The horizontal lines are the observational results by White et al. (1993; dotted line) and by Biviano et al. (1993; dashed line) for clusters with mass larger than the above value.

**Figure 3.** The cluster distribution (heavy dots) superimposed on the DM particle distribution in a slice $10\,h^{-1}$ Mpc thick for a box of side $640\,h^{-1}$ Mpc. The upper panel is for the SCDM model and the lower panel for the OCDM model. Initial random phases are the same for the two models, so that the structures in the plots can be compared directly. It is interesting to note that clusters are strongly correlated with the intersection of filaments of the DM distribution.

**Figure 4.** Comparison between the two-point correlation function for our CHDM cluster simulations (open dots) and for the PM N-body results from Klypin & Rhee (1994) based on the same initial spectrum (filled dots). Their results are obtained with PM simulations having $256^3$ grid points on a box of side $200\,h^{-1}$ Mpc. Our results are the average over 50 realizations, and errors are $1\sigma$ scatter over this ensemble. The KR94 results are an average over 2 realizations and the error bars are quasi-Poissonian estimates.

**Figure 5.** The $J_3(R)$ integral as a function of the scale $R$ for the simulated (open circles) and real Abell/ACO (filled circles) cluster distributions. Error bars are plotted only for the simulations and correspond to $1\sigma$ scatter over the ensemble of 50 realizations.

**Figure 6.** Comparison between the pdf's for real (filled circles) and simulated (open circles) cluster distributions. Results only for the SCDM and CHDM are shown at $R_{sm} = 20$ and $40\,h^{-1}$ Mpc. For the simulations, the analysis is realized in redshift space and the error bars correspond to cosmic r.m.s. scatter.

**Figure 7.** Comparison between the pdf's of simulated cluster distributions and the theoretical models. We plot only results for SCDM and CHDM models at $R_{sm} = 20$ and $40\,h^{-1}$ Mpc. Solid, long-dashed and short-dashed curves correspond to the lognormal, Zel'dovich and Gaussian model, respectively. Error bars are cosmic r.m.s. scatter.

**Figure 8.** Scale dependence of the cluster biasing parameter for the six different models. For reasons of clarity, we plot the corresponding cosmic r.m.s. scatter only for the LOWH model. Similar uncertainties hold also for the other models. The heavy dashed horizontal lines show the range indicated by observational results (see text).

Figure 1

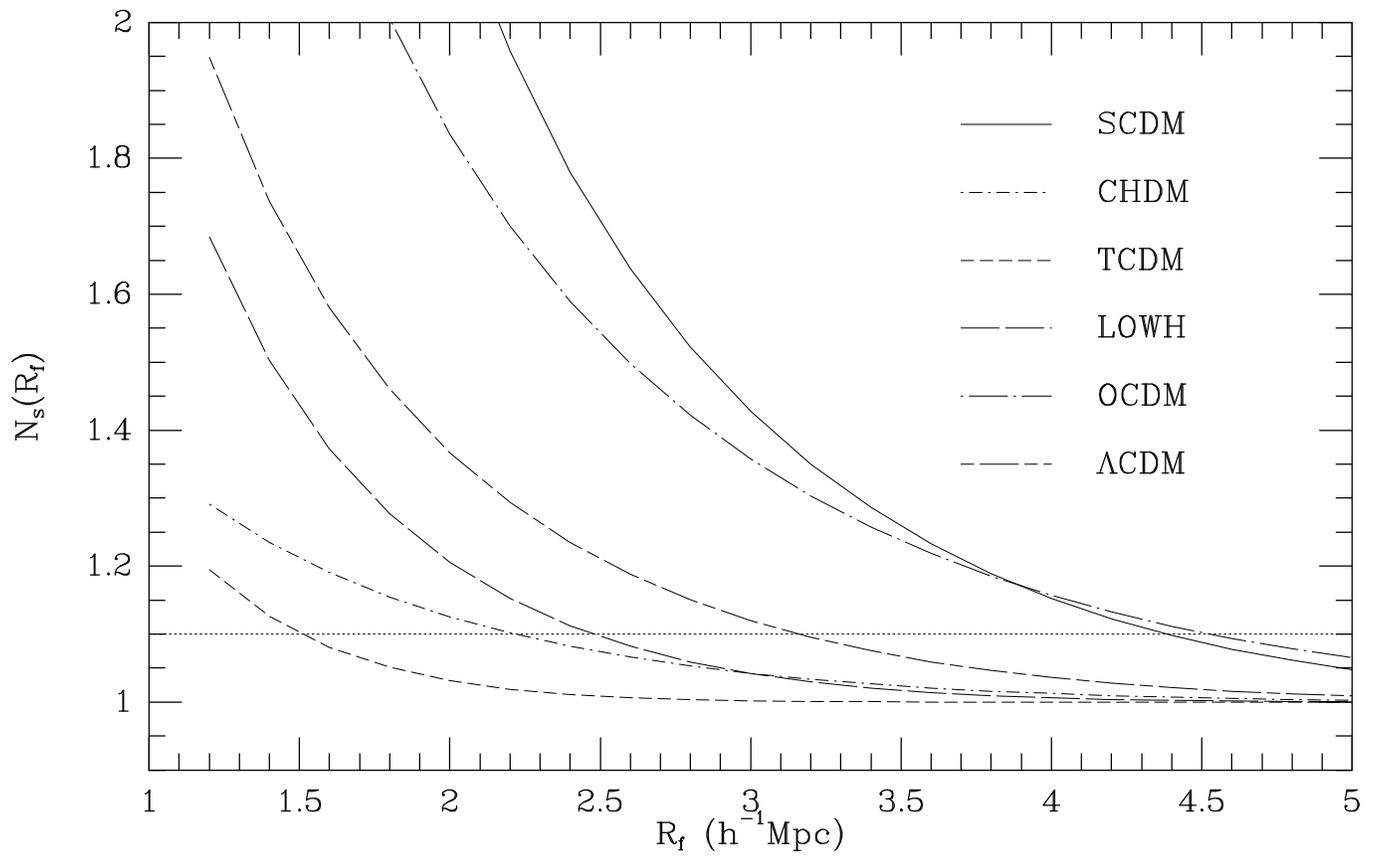

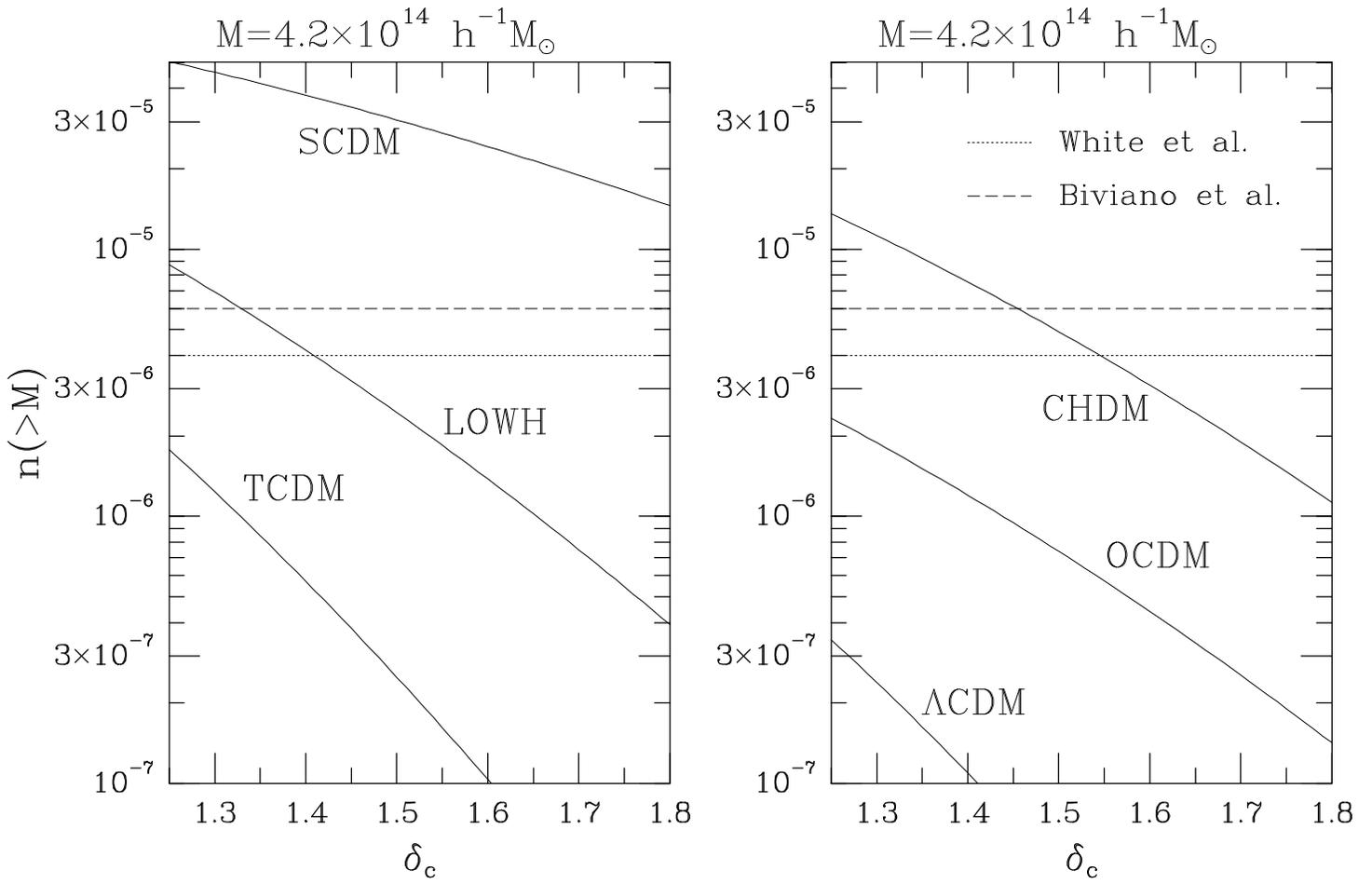

Figure 2

Figure 4

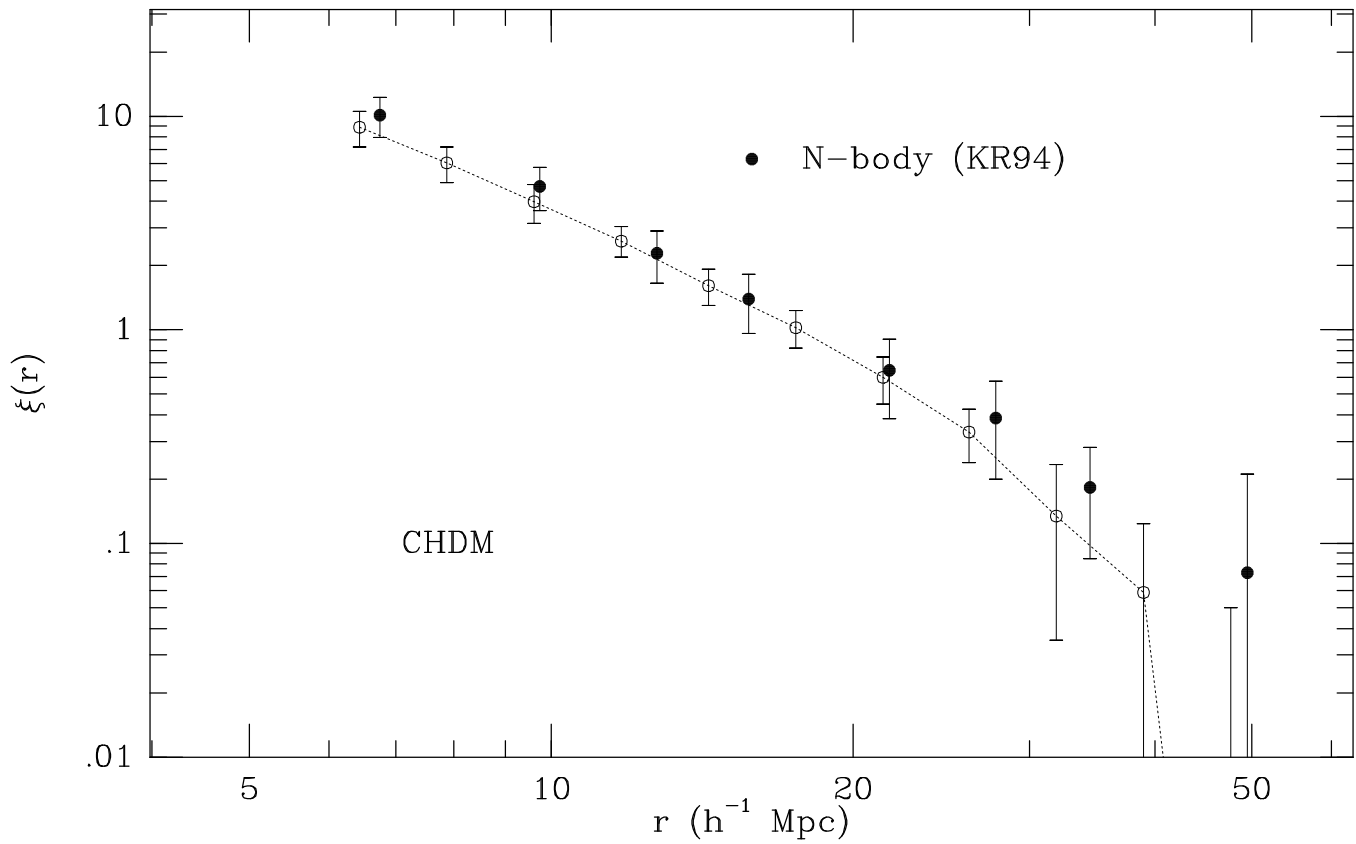

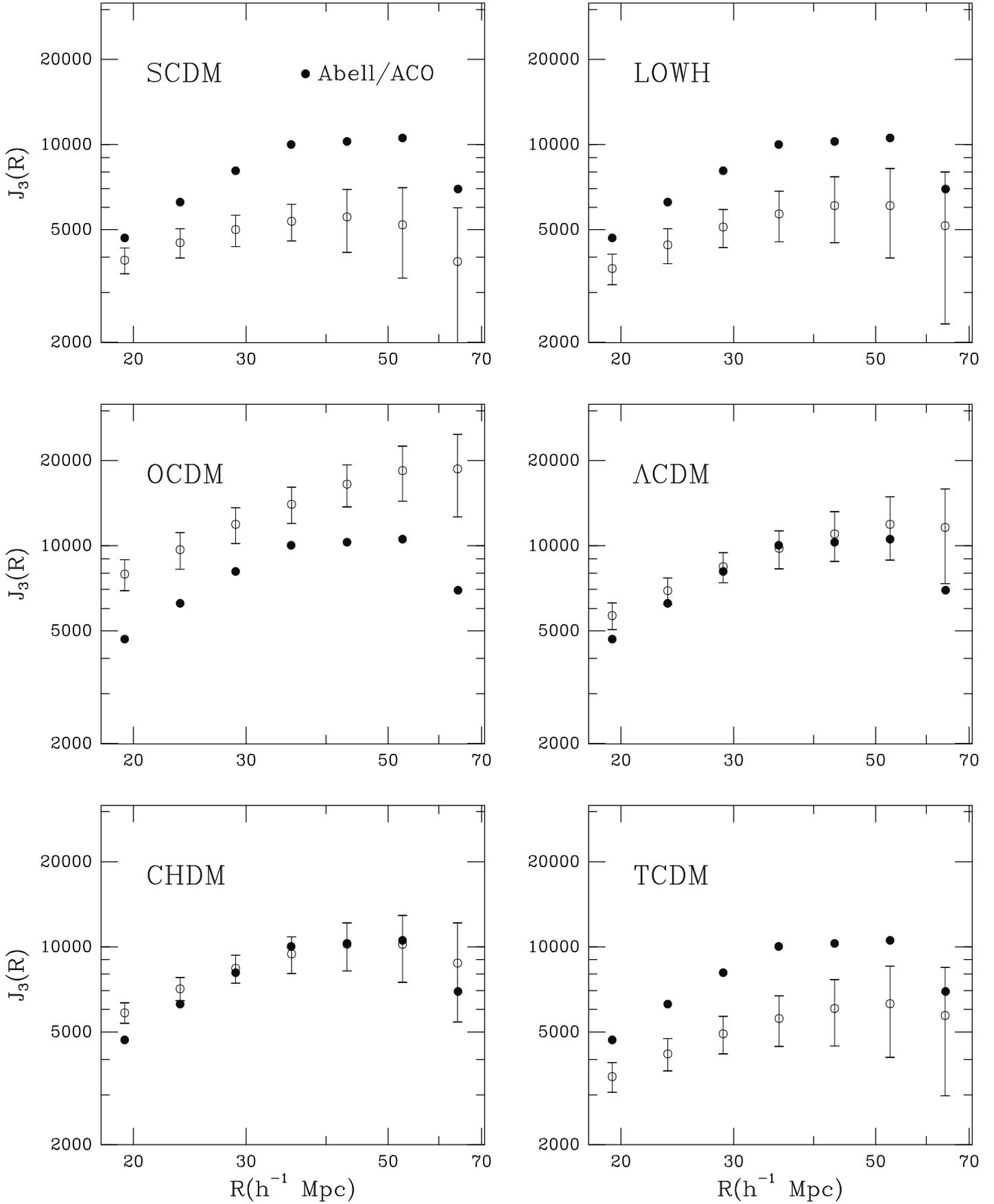

Figure 5

Figure 6

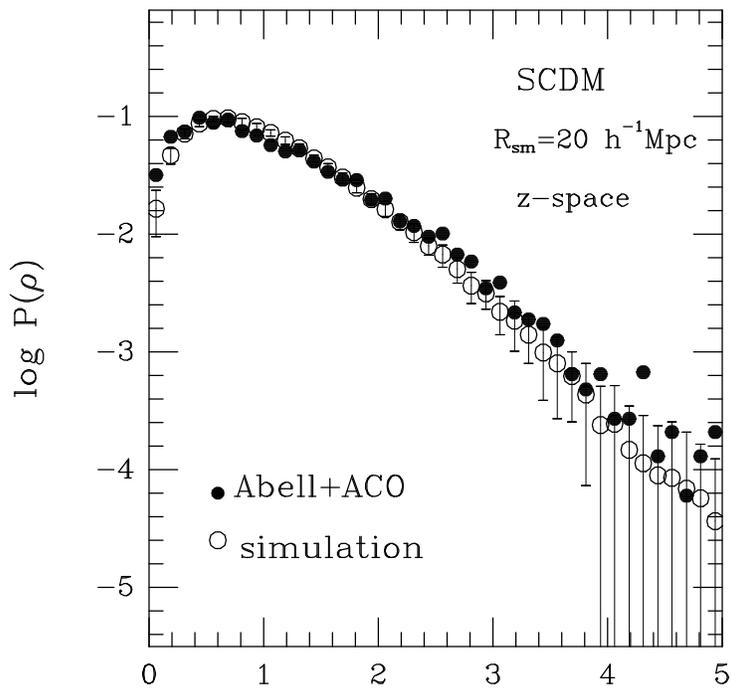
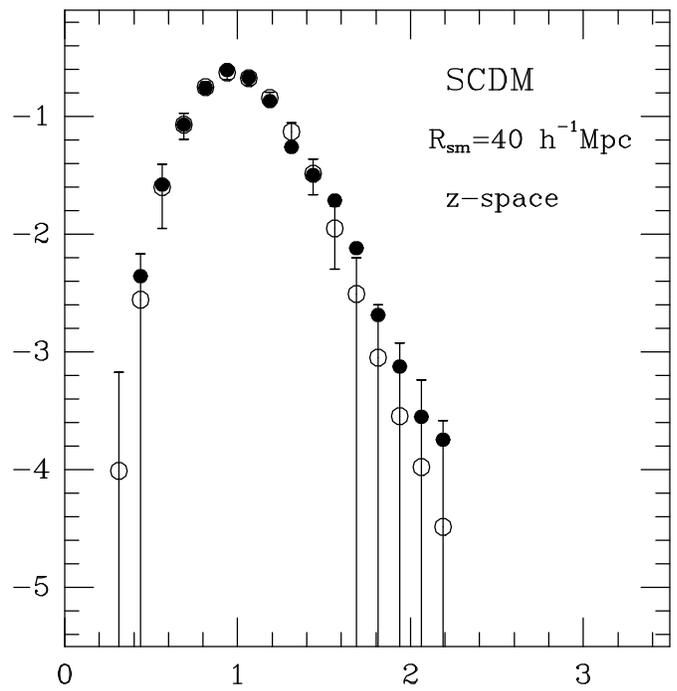
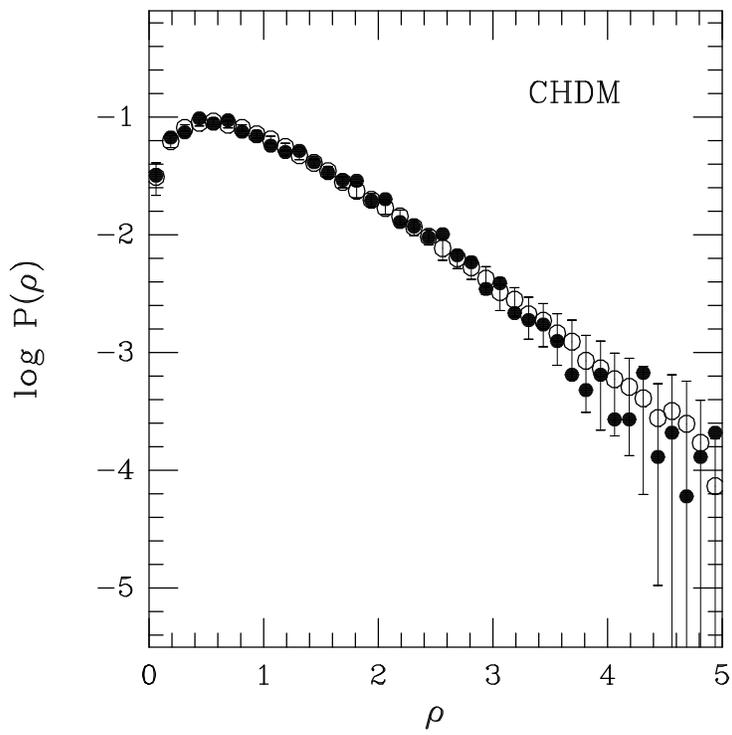
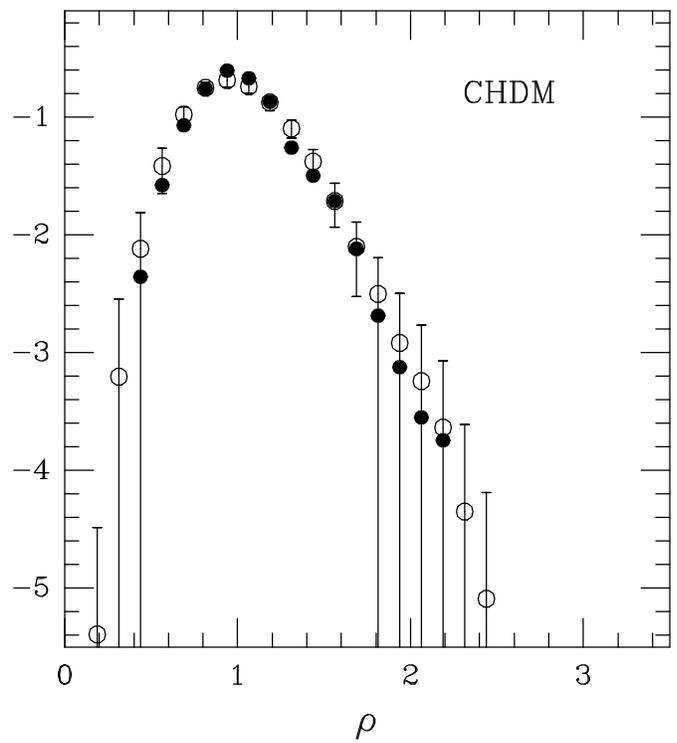

Figure 7

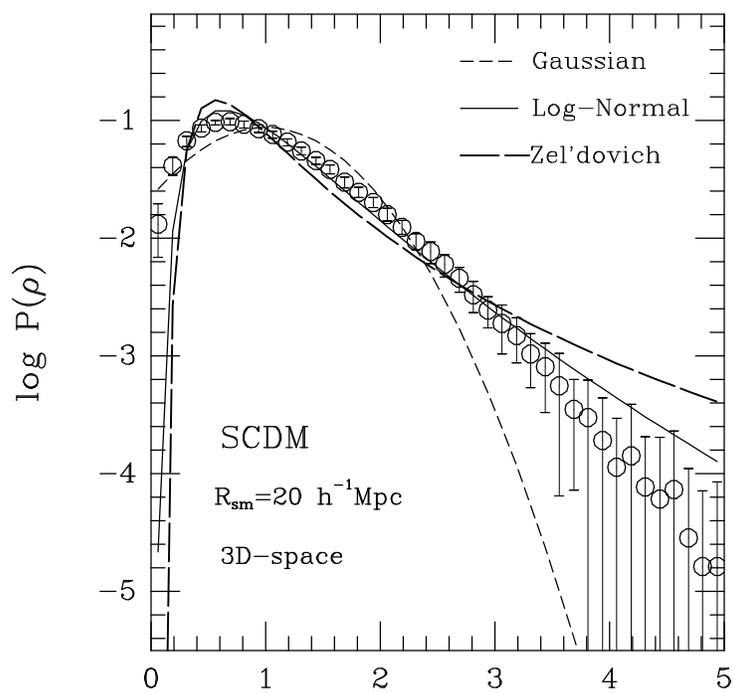
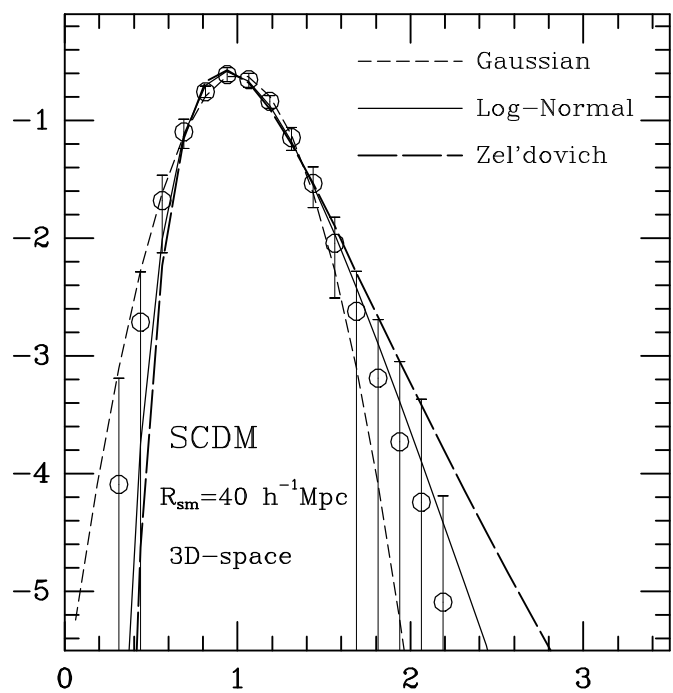
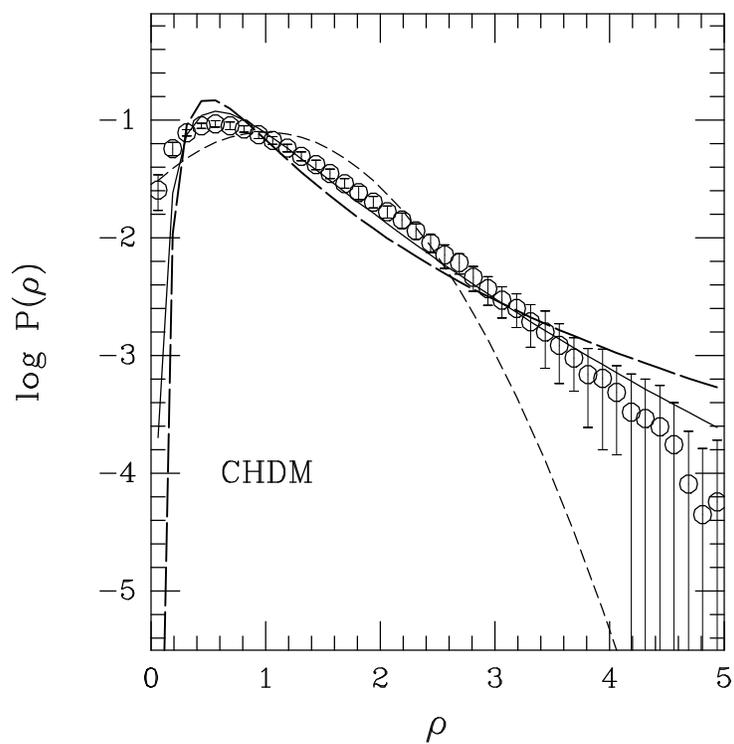
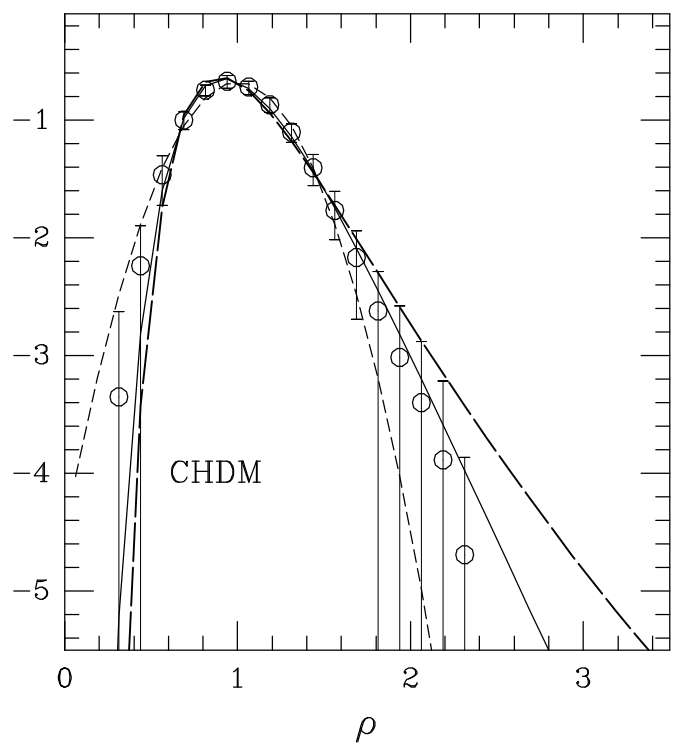

Figure 8

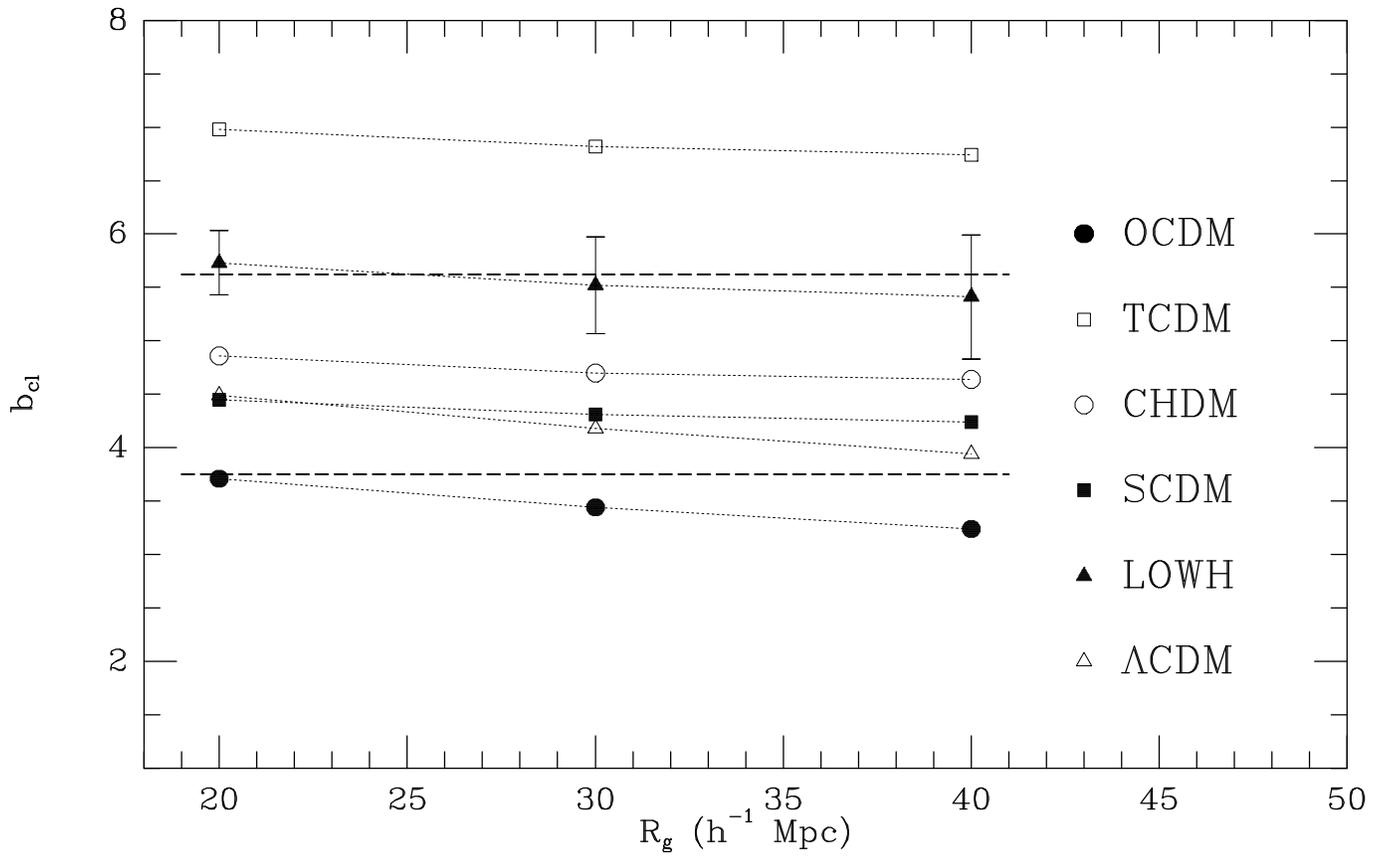